\documentclass[9pt]{sig-alternate}
\usepackage{amsmath,amsfonts,amssymb}
\usepackage{graphicx}
\usepackage{cite}
\usepackage{threeparttable}
\usepackage{booktabs}
\usepackage{multirow}
\usepackage[ruled,linesnumbered,vlined,algoruled]{algorithm2e}
\usepackage{subfig}
\usepackage{bm}
\usepackage{float}
\usepackage{color}
\newcommand{\hu}{}

\newcommand{\bw}{\ensuremath{W_{S\!M\!B}}\xspace}
\newcommand{\dw}{\ensuremath{W_{C\!D}}\xspace}

\begin{document}

\title{Optimizing Memory Efficiency for Convolution Kernels on Kepler GPUs
\vspace{-16pt}}

\author{
Xiaoming Chen, Jianxu Chen, Danny Z. Chen, and Xiaobo Sharon Hu\\
\affaddr{Department of Computer Science and Engineering, University of Notre Dame}\\
\affaddr{Notre Dame, IN 46556, USA}\\
\email{\{xchen7, jchen16, dchen, shu\}@nd.edu}
}

\CopyrightYear{2017}
\setcopyright{acmcopyright}
\conferenceinfo{DAC '17,}{June 18-22, 2017, Austin, TX, USA}
\isbn{978-1-4503-4927-7/17/06}\acmPrice{\$15.00}
\doi{http://dx.doi.org/10.1145/3061639.3062297}

\maketitle

\begin{abstract}
Convolution is a fundamental operation in many applications, such as computer vision, natural language processing, image processing, etc. Recent successes of convolutional neural networks in various deep learning applications put even higher demand on fast convolution. The high computation throughput and memory bandwidth of graphics processing units (GPUs) make GPUs a natural choice for accelerating convolution operations. However, maximally exploiting the available memory bandwidth of GPUs for convolution is a challenging task. This paper introduces a general model to address the mismatch between the memory bank width of GPUs and computation data width of threads. Based on this model, we develop two convolution kernels, one for the general case and the other for a special case with one input channel. By carefully optimizing memory access patterns and computation patterns, we design a communication-optimized kernel for the special case and a communication-reduced kernel for the general case. Experimental data based on implementations on Kepler GPUs show that our kernels achieve 5.16$\times$ and 35.5\% average performance improvement over the latest cuDNN library, for the special case and the general case, respectively.
\end{abstract}


\vspace{-10pt}

\vspace{-12pt}

\section{Introduction}
Convolution is a fundamental operation in many image processing and computer vision applications. For example, image convolution is a key component in numerous basic image processing routines, such as edge detection~\cite{image_book2007}, smoothing~\cite{image_book2007}, template-based object detection~\cite{image_tmi1989}, etc. Recently, convolutional neural networks (CNNs)~\cite{cnn_1989} have become a powerful deep learning model which has been widely adopted in various computer vision applications, such as image recognition~\cite{cnn_2014}, image classification~\cite{cnn_nips2012}, object detection~\cite{cnn_cvpr2014}, etc. State-of-the-art CNNs typically have quite a few convolutional layers.
Propagating through these convolutional layers is always a computation bottleneck in both the training and inference phases of CNNs. 



With the rapid development of many-core parallel processors, new methods have been developed by leveraging high computation throughput and memory bandwidth of graphics processing units (GPUs) to accelerate convolution operations. These methods can be roughly classified into four categories: (1) general matrix multiplication (GEMM) based convolution~\cite{gemm_2006,cudnn_2014}, (2) direct convolution~\cite{direct_hpcc2015,direct_2014,convnet2}, (3) fast Fourier transform (FFT) based convolution~\cite{fft_2013,fft_2014,fft_2016}, and (4) the Winograd algorithm~\cite{winograd_cvpr2016,winograd_codes2016}.

Convolution can be easily converted into a multiplication of two matrices by unrolling all the involved convolution operations~\cite{gemm_2006}. Highly optimized GEMM kernels (e.g., cuBLAS~\cite{cublas_web}) can be invoked to compute matrix multiplications. This is the default method in Caffe~\cite{caffe}, a popular deep learning framework. Although good performance can be attained, it requires a huge amount of additional memory.
Recently, cuDNN~\cite{cudnn_2014} adopted a GEMM-like method, in which sub-blocks of the input matrices are constructed in on-chip memory at run-time, and thus no additional memory is needed. A direct method was proposed in~\cite{direct_hpcc2015}, but the reported performance is not good enough when there are more than 100 channels. In~\cite{direct_2014}, optimization techniques were discussed for direct convolution on GPUs, but the proposed method was not compared with any public library. Cuda-convnet2~\cite{convnet2} also implemented direct convolution on GPUs, but there is no detailed document to describe its methodology or performance. FFT-based convolution~\cite{fft_2013,fft_2014,fft_2016} can reduce the arithmetic complexity compared with direct methods. However, the filters need to be padded to the same size as the input image, which incurs additional memory and computation time. In addition, in order to reuse the Fourier transform of the filters, the batch size should be big enough. Recent studies have shown that the Winograd algorithm can significantly reduce the arithmetic complexity for the 3$\times$3 filter~\cite{winograd_cvpr2016,winograd_codes2016}, at the cost of increased memory usage and filter size dependent specialized processing.


Although FFT-based methods and the Winograd algorithm can be faster than direct methods in some cases, they are not universal. Direct convolution is still fundamental and considered the best in general. In this paper, we aim to improve the memory efficiency of direct convolution on GPUs, targeting at two cases: (1) a special case with one input channel, which appears in  numerous image processing applications and the input layer of CNNs (for grayscale images), and (2) the general case for CNNs. Specifically, we introduce a general model to address the mismatch between the shared memory bank width of GPUs and computation data width of threads. Based on this model, by carefully optimizing the memory access patterns and computation patterns, we design a communication-optimized kernel for the special case and a communication-reduced kernel for the general case. Experimental data based on implementations on Kepler GPUs show that our convolution kernels achieve 5.16$\times$ and 35.5\% average performance improvement compared with the latest cuDNN library, for the special case and the general case, respectively.



\section{Problem Formulation}\label{sec:model}

In this section, we present the problem formulation, which illustrates the main challenges and the general model that we propose to overcome such challenges.

\subsection{GPU Memory Constraints and Modeling}\label{sec:bank}
Most GPU programs are memory bandwidth hungry. GPUs 
{\hu usually have a complex memory hierarchy} subject to different constraints. Global memory (GM) accesses should be coalesced in order to reduce latency. Bank conflict should be avoided when accessing the shared memory (SM). For the constant memory (CM), {\hu all the access addresses} within one warp should be identical to take full advantage of the broadcast mechanism. These are basic constraints that GPU programs should satisfy to achieve good performance.

\begin{figure}[b]
  \centering
  \includegraphics[width=.8\columnwidth]{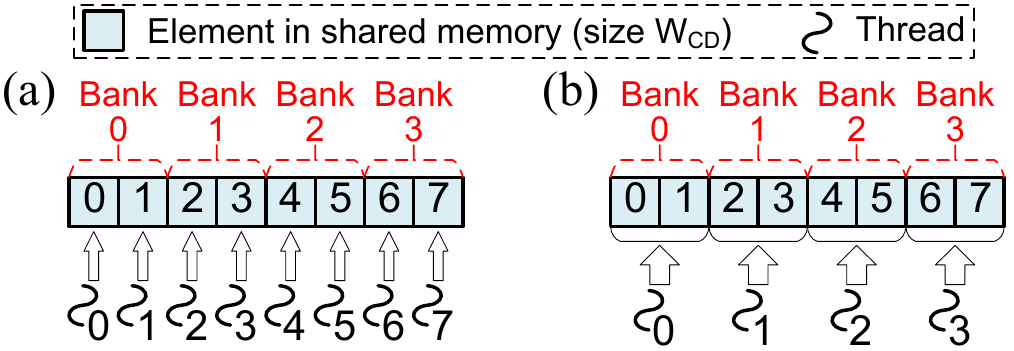}\\
  \vspace{-8pt}
  \caption{Different SM access patterns. (a) Conventional approach. (b) Matched approach.}\label{fig:shmem}
\end{figure}

The SM bank width, which also plays an important role in GPU performance, however, has received less attention from programmers and researchers. We elaborate this problem below. Let \bw be the \textit{SM bank width}. (\bw is 8 (bytes) on Kepler and 4 on other GPU architectures.) Further, let \dw be the \textit{computation data width} for each thread. For example, if each thread takes {\tt float} as the minimum unit fort computation, then $\dw=4$. The relation between \bw and \dw can be described by
 \vspace{-4pt}
\begin{equation}\label{eq:model}
  \bw =n \cdot \dw.
   \vspace{-4pt}
\end{equation}
If $n\!=\!1$, the SM bank width and computation data width are {\it matched}; otherwise, they are {\it unmatched}. Mismatch between \bw and \dw frequently occurs in practice. Even when \bw is 4, \dw can be 2 (for {\tt short} or {\tt fp16}) or 1 (for {\tt char}). Fig.~\ref{fig:shmem} illustrates the impact of a mismatch. Consider multiple threads reading from or writing to the SM. A conventional method shown in Fig.~\ref{fig:shmem}a is often used, where contiguous threads access contiguous elements, as it is easy to program. But, such a method may fail to fully utilize the available SM bandwidth. For example, if $n\!=\!2$, as shown in Fig.~\ref{fig:shmem}a, any two accesses that fall into the same bank have to be serialized.  Yet, if we can double the computation data width {\hu through intelligent thread layout and computation pattern redesign} so that $\bw\! =\! \dw$, as in Fig.~\ref{fig:shmem}b, then each thread can obtain 2 elements together in a single access, resulting in a 2$\times$ improvement in the SM bandwidth.

\begin{figure}[t]
  \centering
  \includegraphics[width=.65\columnwidth]{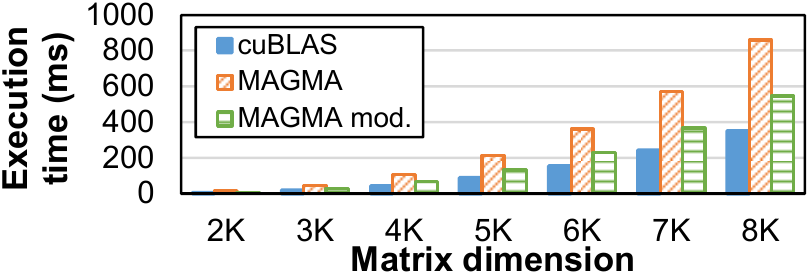}\\
  \vspace{-8pt}
  \caption{Execution time comparison for the single-precision GEMM on a Kepler K40m GPU.}\label{fig:magma}
  \vspace{-8pt}
\end{figure}

To demonstrate the importance of this problem, we compare the performance of the single-precision GEMM on a Kepler K40m GPU, as shown in Fig.~\ref{fig:magma}. MAGMA is highly optimized for Fermi and is faster than cuBLAS on the Fermi architecture~\cite{magma_jhpca2010}; however, it becomes 2.4$\times$ slower than cuBLAS on the Kepler architecture. The SM bank width of the Kepler architecture is twice of that of Fermi, causing a mismatch between \dw and \bw for the MAGMA kernel that operates on {\tt float}, which results in the loss of half of the SM bandwidth. Yet, a modification to the MAGMA kernel by matching \dw with \bw saves 36\% of the execution time on average.

Consequently, for applications that are sensitive to the SM bandwidth, memory access patterns and computation patterns should be reorganized to match \dw with \bw. That is, each thread should be designed such that it accesses and computes $n$ basic elements as a single unit. In this way, we can obtain an $n\times$ improvement in the SM bandwidth.

\subsection{Data Sharing in Convolution}

\begin{figure}[t]
  \centering
  \subfloat[]
  {
  \label{fig:conv}
  \includegraphics[width=.7\columnwidth]{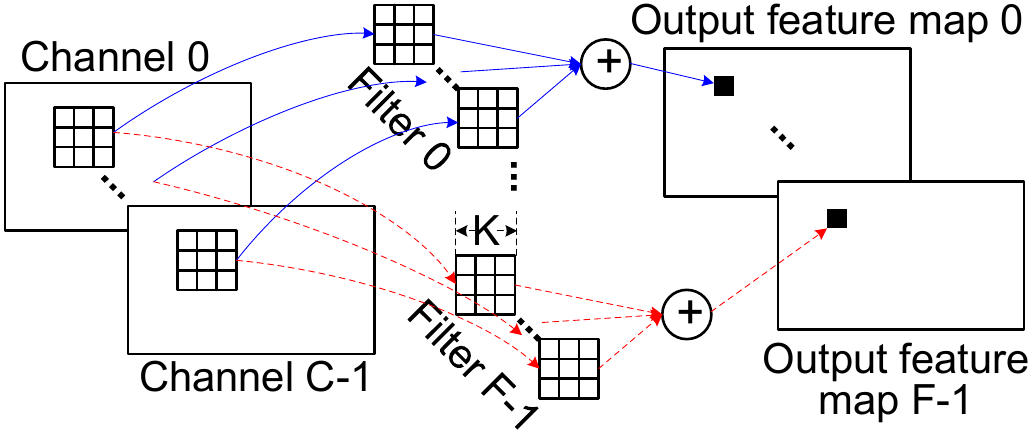}
  }
  \subfloat[]
  {
  \label{fig:reuse}
  \includegraphics[width=.25\columnwidth]{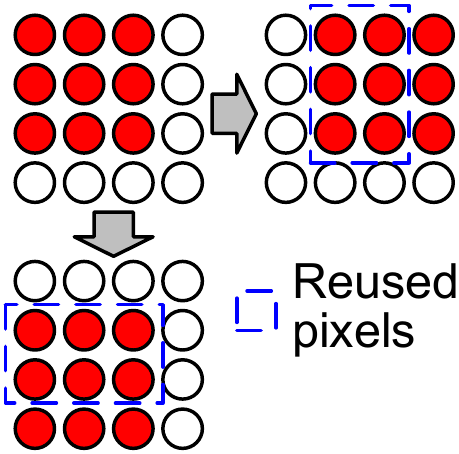}
  }
  \vspace{-8pt}
  \caption{Basics of convolution. (a) A general convolution operation in CNN ($C$ is the \# of channels, $K$ is the filter size, and $F$ is the \# of filters). (b) Data reuse in convolution (the solid circles mark pixels that are being used for convolution).}\label{fig:basic}
  \vspace{-8pt}
\end{figure}

One key aspect in developing highly efficient convolution operations on GPUs is to maximize data sharing, which is also a key factor for communication reduction. Consider the general case of convolution {\hu in CNNs} (see Fig.~\ref{fig:conv}). 
Fig.~\ref{fig:reuse} illustrates a simple data reuse method, in which pixels can be reused in both the horizontal and vertical directions as indicated by the solid circles within the dashed boxes. A simple  analysis shows that an input pixel can be used up to $K\!\times\! K\!\times\! F$ times, where $K$ is the filter size and $F$ is the number of filters. This feature should be fully exploited to reduce both GM and SM accesses. For this aim, elaborate memory access patterns and computation patterns need to be used, while still satisfying the basic constraints of the GPU memory hierarchy.

Another challenge arises from the SM bank width model presented in the previous subsection. As convolution is SM bandwidth bounded, when designing convolution kernels on GPUs, we must strive to match \dw and \bw, in order to fully utilize the SM bandwidth.

Taking all these requirements into consideration, the goal of this paper is to develop general solutions for convolution on GPUs, such that (i) the memory communication is reduced as much as possible, (ii) the basic constraints of the GPU memory hierarchy are satisfied, and (iii) the SM bank width and the computation data width are matched. 

\section{Convolution for Special Case}\label{sec:special}
This section presents our convolution kernel for the special case, in which the input has only one channel ($C\!=\!1$ in Fig.~\ref{fig:conv}). This case arises at the first layer of CNNs (for grayscale images) and in many image processing applications. We first show how we design the thread layout and then discuss how we achieve optimal memory accesses.

\subsection{Thread Layout}\label{sec:threads}
The goal of thread layout is to judiciously allocate computation to thread blocks (TBs) and individual threads to maximize both coarse-grained and fine-grained parallelism.
Fig.~\ref{fig:reuse} depicts the general concept for our parallelization methodology, in which each thread keeps $K\!\times\! K$ pixels {\hu of the input image} used by a convolution operation in the register. However, directly applying this method would cause a problem. A thread cannot move right and down simultaneously. In other words, once a thread moves to the right to compute the next convolution, it loses some pixels needed by the convolution below. One could use additional registers to store such lost pixels, but a lot more registers would be required. To resolve this issue, we propose an alternative scheme to maximize both parallelism and 2D data sharing.

\begin{figure}[t]
  \centering
  \includegraphics[width=.38\columnwidth]{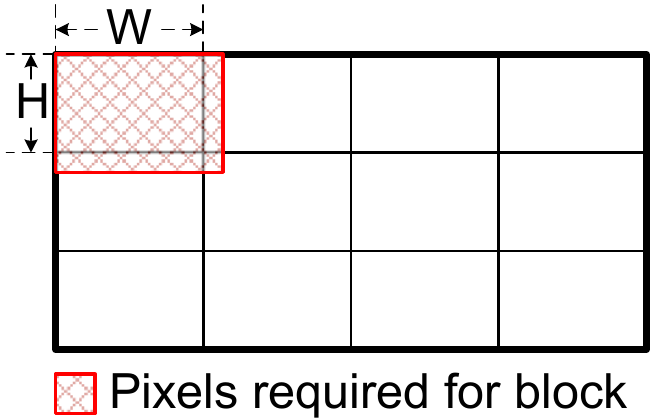}\\
  \vspace{-8pt}
  \caption{Image partitioning.}\label{fig:block}
  \vspace{-16pt}
\end{figure}

To achieve coarse-grained parallelism, we partition the input image into blocks of size $H\!\times\! W$ each (see Fig.~\ref{fig:block}). Such partitioning enables data sharing along the vertical direction, since {\hu one row of the input image} can be used by the convolutions of $K$ rows. A TB with $W$ threads handles one image block. Different image blocks are assigned to different TBs so that they are computed in parallel. Each block needs some additional pixels outside its right and bottom boundaries to compute convolutions.

\begin{figure}[t]
  \centering
  \subfloat[]
  {\label{fig:speca}
  \includegraphics[width=.42\columnwidth]{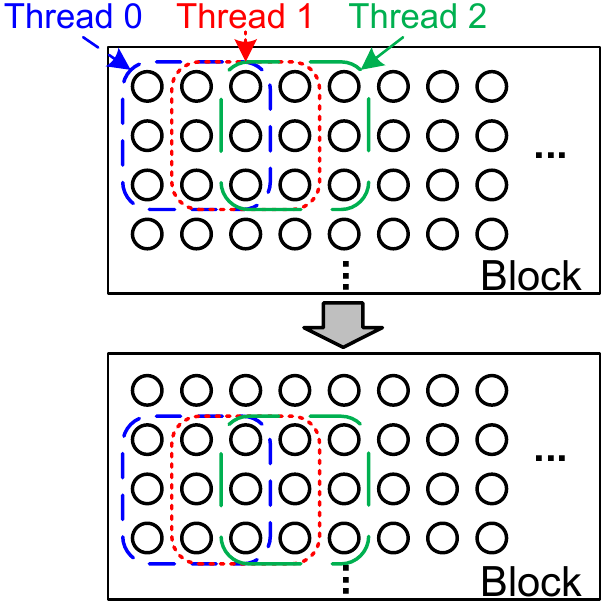}
  }
  \subfloat[]
  {\label{fig:specb}
  \includegraphics[width=.53\columnwidth]{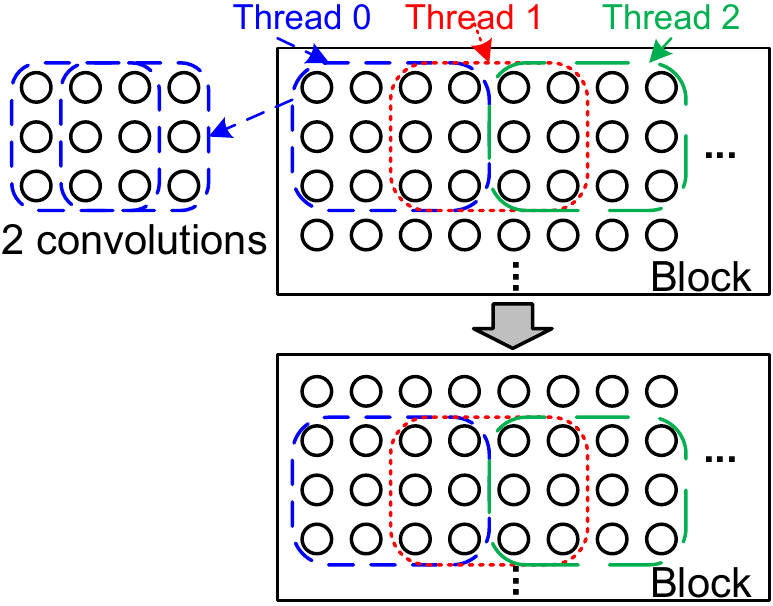}
  }
  \vspace{-8pt}
  \caption{Our convolution method for the special case ($K=3$ in this example). (a) For matched \dw and \bw. (b) For unmatched \dw and \bw ($n=2$ in this illustration).}\label{fig:specconv}
  \vspace{-8pt}
\end{figure}

In terms of fine-grained parallelism, all the $W$ threads in a TB compute convolutions of one row in parallel, as illustrated in Fig.~\ref{fig:speca}. Once a row is finished, the $W$ threads move down to compute the next row. Thus we read a new row from the input image for each down movement. This process continues until reaching the bottom of the block.

\subsection{Optimizing Memory Accesses}

We now discuss how we schedule memory accesses in coordination with the thread layout design in Section~\ref{sec:threads}. Since the filters in the special case (with only 1 input channel) are typically  small,  they can reside in the CM and no further scheduling is needed. We focus our discussion on accessing the input image in the GM to minimize GM communication.

We first consider the simple case of $\dw \!= \!\bw$. 
For each row of the block, we first read it into the SM (including the needed pixels outside the block boundaries), and then the $W$ threads read their corresponding pixels into respective registers. This process allows horizontal data sharing and avoids redundant reads from the GM as adjacent threads share some common pixels.
Hence, our 2D data sharing method works as follows: in the horizontal direction, the SM provides inter-thread data sharing; in the vertical direction, intra-thread data sharing is achieved through the private registers of the threads.

A simple analysis shows that each pixel in a block is read from the GM only once, which is, of course, the theoretical lower bound. For the entire image, only those pixels which are needed by a block and outside the block boundaries are read more than once. But, the proportion of such halo pixels is small. As a result, this method is (almost) communication-optimal for GM accesses.

When considering the SM bank width model presented in Section~\ref{sec:bank}, the above thread layout and memory access schedule are suitable only when $n\!=\!1$, i.e., $\dw \!=\! \bw$. On the Kepler architecture, $n\!=\!2$ if we use {\tt float} as the basic computation unit. Following the general idea depicted in Fig.~\ref{fig:shmem}b,  we propose to have each thread read, write, and compute $n$ pixels together (using built-in data types such as {\tt float2} or {\tt float4}). Each thread is responsible for $n$ contiguous output pixels in each row and $n\!\times\! H$ output pixels in the block (for one output feature map). The number of threads in a TB is reduced to $\frac{W}{n}$. With this approach, each thread needs a few more registers ($O(K \!\times \!(n\!-\!1))$)  to store $K\!\times\! (K\!+\!n\!-\!1)$ pixels that are used by the convolutions for $n$ contiguous output pixels. Fig.~\ref{fig:specb} shows our modified convolution method specifically for the Kepler architecture.

\subsection{Implementation}

Algorithm~\ref{alg:spec} outlines the flow of our special case convolution method at the TB level. The algorithm starts by reading $K$ image rows of the block into the SM (line 1). After that, the first $K\!-\!1$ rows are read into the threads' registers (line 3). Then {\hu convolution is performed on all the rows within the block iteratively in a loop} (lines 4-11). For each row, the data are first read from the SM into the threads' registers (line 6), and then each thread computes convolutions for all the filters (lines 7 and 8). We use a prefetching mechanism to overlap computations and GM accesses. Before the threads' computation tasks, the next image row of the block is prefetched into the threads' registers (line 5). Although this operation may take a long time, it can be overlapped with convolution computations, since they have no data dependency. After prefetching is finished, the prefetched data are written into the SM (line 10).

\begin{algorithm}[t]
\scriptsize
\caption{\small Convolution for the special case on GPUs.}\label{alg:spec}

Load rows 0 to $K\!-\!1$ (the first $K$ rows) of the block into the SM;

{\tt \_\_syncthreads()};

Each thread loads $(K\!-\!1)\!\times \!(K\!+\!n\!-\!1)$ pixels from the SM into register;

\For {$k\!=\!K\!-\!1$; $k\!<\!H\!+\!K\!-\!1$; $\!+\!+\!k$}
{
    Prefetch row $k\!+\!1$ of the block into register;

    Each thread loads the latest row from the SM into register;

    \For {$f\!=\!0$; $f\!<\!F$; $\!+\!+\!f$}
    {
        Each thread computes $n$ convolutions for filter $f$ and writes the results back to the GM;
    }

    {\tt \_\_syncthreads()};

    Store the prefetched row into the SM;

    {\tt \_\_syncthreads()};
}
\end{algorithm}

The above algorithm is quite memory efficient. When computing convolutions, the involved pixels are in registers so the latency can be ignored. The filters are fetched from the CM. In our method, all the threads within a warp always compute convolutions using the same filter at the same time, so they always access the identical address, which is the best case for the CM. As the filters are quite small in the special case, a high hit rate of the constant cache can be expected. When accessing the SM and GM, contiguous threads always read or write contiguous addresses (at the granularity of $n$ pixels as a single unit), so both coalesced GM access and conflict-free SM access are achieved. Our experimental results in Section~\ref{sec:result} support our analysis here.

\section{Convolution for General Case}\label{sec:general}

This section presents our convolution kernel for the general case, where the input has multiple channels (see Fig.~\ref{fig:conv}). Note that the method for the special case cannot be applied here for the following reasons. For the special case, we can keep the needed pixels in the threads' registers since the filters are small. Thus, we can finish one convolution at once with $K\!\times\! K$ fused multiply-add (FMA) operations. With multiple channels, the involved pixels of one convolution  cannot entirely reside in the registers. Hence, the computation of one convolution must be divided into multiple steps, and the intermediate results should be accumulated in the registers. In addition, the filters (proportional to the number of channels) become larger and may no longer fit in the CM. Instead, the GM needs to store both the filters and the input image. Our basic idea for the general case  is inspired by the blocked GEMM method for GPUs~\cite{magma_jhpca2010}, but we optimize memory communication by maximally sharing data.

\subsection{Thread Layout}
Similar to the special case, each input channel of the input image is partitioned into blocks of size $H\!\times \!W$ each (see Fig.~\ref{fig:block}). We use a 2D TB layout which is similar to that adopted by the blocked GEMM method~\cite{magma_jhpca2010}. Since in the general case, a TB cannot be responsible for all the filters, we divide the computation into a 2D TB layout of size $TB_{X}\!\times \!TB_{Y}$. In the X dimension, a TB is responsible for $F_{TB}$ contiguous filters, where $TB_{X}=\left\lceil{F}/{F_{TB}} \right\rceil$. In the Y dimension, a TB is responsible for $C$ image blocks at the same location of all the $C$ channels. Within a TB, we use a 2D thread layout of size $T_X\!\times\! T_Y$. Each thread is responsible for $W_T$ output pixels and $F_T$ filters, where $T_X=\frac{F_{TB}}{F_T}$ and $T_Y=\frac{W\!\times\! H}{W_T}$. Each thread keeps $F_T\!\times\! W_T$ pixels in the register to store the intermediate convolution results for the $W_T$ pixels and $F_T$ filters. In the following subsections, our discussion will focus on optimizing memory access patterns to reduce memory communication.

\subsection{Optimizing Memory Accesses}

\begin{figure}
  \centering
  \includegraphics[width=.98\columnwidth]{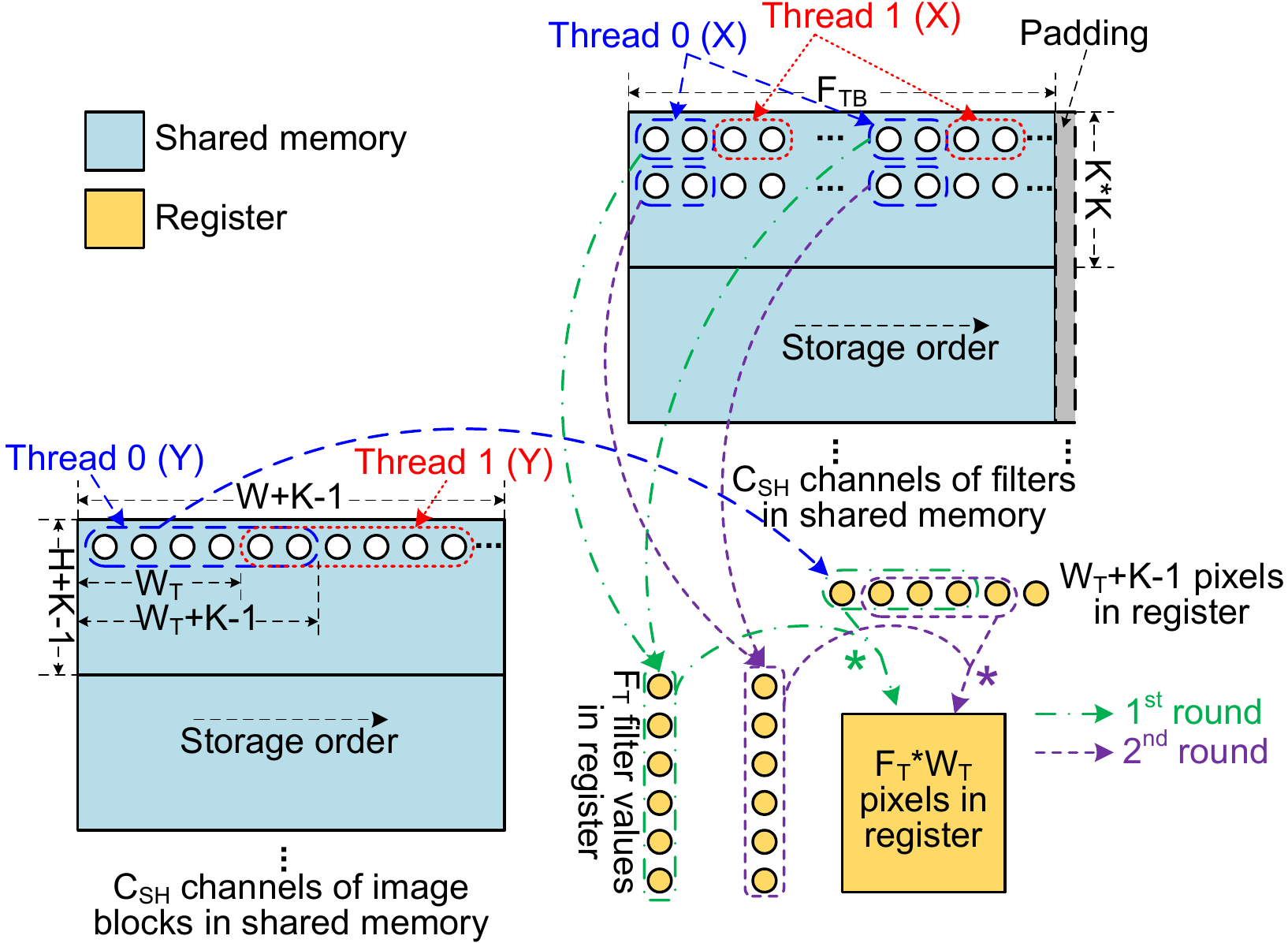}\\
  \vspace{-8pt}
  \caption{Our convolution method for the general case ($n\!=\!2$ in this illustration).}\label{fig:genrconv}
  \vspace{-8pt}
\end{figure}

Fig.~\ref{fig:genrconv} depicts our convolution method for the general case at the TB level. To improve the GM efficiency, we store $C_{SH}$ channels of image blocks (including the needed halo pixels outside the block) and the filters in the SM (shown as the two blue boxes in Fig.~\ref{fig:genrconv}). When reading filter values from the GM to the SM, since the block is transposed, padding (the gray box in Fig.~\ref{fig:genrconv}) is required for the SM to avoid bank conflict. For the image blocks, pixels are directly read into the SM without transposition, so padding is not needed.

To increase data sharing within a thread, the $W_T$ output pixels computed by one thread are contiguous along the horizontal direction. This is a major difference from the blocked GEMM method~\cite{magma_jhpca2010} where contiguous output pixels are computed by contiguous threads. Computing $W_T$ contiguous output pixels by one thread involves reading $(W_T\!+\!K\!-\!1)\times \!K\!\times\! C$ pixels from the SM, instead of $W_T \!\times \!K\!\times \!K\!\times\! C$ if they are computed by different threads. As $C$ may be large, it is impossible to put all of these involved pixels in the register. So we only keep a row of $W_T\!+\!K\!-\!1$ input pixels in each thread's register, and the convolution results are accumulated iteratively. The $W_T\!+\!K\!-\!1$ input pixels are used in $K$ rounds of computation of $W_T$ output pixels. A round of computation refers to an FMA operation in a convolution.

When reading filter values from the SM, we use the same thread layout as that used in the blocked GEMM method~\cite{magma_jhpca2010}. However, in order to meet the requirement of the SM bank width model, each thread should read $n$ contiguous values as a single unit along the horizontal direction, as illustrated in Fig.~\ref{fig:genrconv} (the upper blue box). This method is conflict-free as contiguous threads in the X dimension read contiguous units from the SM.

Our thread layout and memory access patterns also avoid bank conflict when accessing the SM for image blocks. As the X dimension of the 2D thread layout is assigned along the feature direction, $T_X$ contiguous threads in the X dimension access the identical address of the SM for image blocks, which benefits from the broadcast mechanism of the SM.

The only issue of this approach is at the writing back phase. Writing the results back to the GM is not coalesced, as contiguous threads in the X dimension compute different output feature maps. However, we have found that in the general case convolution, the writing back phase consumes very little time, so we do not optimize the uncoalesced writing back operations. If one wants to make the writing back coalesced, the SM can be used as a buffer to reorganize the data layout. However, this would lead to additional cost including the SM latency and TB barriers.

\subsection{Implementation}

\begin{algorithm}[t]
\scriptsize
\caption{\small Convolution for the general case on GPUs.}\label{alg:genr}

Register: $rAcc[F_T][W_T]$, $rImg[W_T\!+\!K\!-\!1]$, $rFlt[F_T]$;

Shared memory: $shImg[C_{SH}][H\!+\!K\!-\!1][W\!+\!K\!-\!1]$, $shFlt[C_{SH}][K\times K][F_{TB}\!+\!padding]$;

Clear $rAcc$;

Load $C_{SH}$ channels of image blocks into $shImg$;

Load $C_{SH}$ channels of filters into $shFlt$;

{\tt \_\_syncthreads()};

\For {$c\!=\!0$; $c\!<\!C$; $c+\!=\!C_{SH}$}
{
    Prefetch next $C_{SH}$ channels of image blocks into register;

    Prefetch next $C_{SH}$ channels of filters into register;

    \For {$i\!=\!0$; $i\!<\!C_{SH}$; $\!+\!+\!i$}
    {
        \For {$j\!=\!0;$ $j\!<\!K$; $\!+\!+j$}
        {
            Each thread loads $W_T\!+\!K\!-\!1$ pixels into $rImg$;

            \For {$k\!=\!0$; $k\!<\!K$; $\!+\!+\!k$}
            {
                Each thread loads $F_T$ filter values into $rFlt$;

                $rAcc[0,\cdots,F_T\!-\!1][0,\cdots,W_T\!-\!1]+=rFlt[0,\cdots,F_T\!-\!1]\times rImg[k,\cdots,W_T\!+\!k\!-\!1]$;
            }
        }
    }

    {\tt \_\_syncthreads()};

    Store prefetched image blocks into $shImg$;

    Store prefetched filters into $shFlt$;

    {\tt \_\_syncthreads()};
}

Write $rAcc$ back to the GM;

\end{algorithm}

Algorithm~\ref{alg:genr} outlines our general case convolution method at the TB level. The algorithm starts by clearing the results (line 3) and loading the first $C_{SH}$ channels of image blocks and filters into the SM (lines 4 and 5). After that, a loop iteratively accumulates the results for all the channels (lines 7--19). For each channel, a thread needs to conduct $K$ rows of computations (lines 11--15), and for each row, $K$ rounds of computation are conducted (lines 13--15). The image data are loaded into each thread's register only for each row (line 12), and these data are used by $K$ rounds of computation. The filter data are loaded into the register in each round (line 14). We also use a prefetching method to overlap GM accesses and computation (lines 8, 9, 17 and 18). After the intermediate results of all the channels are accumulated, the final results are written back to the GM (line 20).

Fig.~\ref{fig:genrconv} illustrates the first two rounds of computation for thread $(0,0)$. The thread first loads a row of $W_T\!+\!K\!-\!1$ pixels from the SM into the register. In the first round, it loads $F_T$ filter values from the first row of the SM storing the filters, and then updates the intermediate results by multiplying the $F_T$ filter values and the first $W_T$ pixels (the green dashed lines). In the second round, the $F_T$ filter values are loaded from the second row of the SM, but the pixels are not loaded again, as they are already in the row of $W_T\!+\!K\!-\!1$ pixels with an offset (the purple dashed lines).

Compared with direct GEMM-based convolution methods, our method reduces GM communication by approximately $\frac{1}{K}$, since one image row is used by the convolution of $K$ rows. The SM communication for fetching image pixels is reduced by $\frac{W_T+K-1}{W_T\cdot K}$.


\section{Experimental Results}\label{sec:result}

We have implemented our proposed methods and conducted experiments on a Kepler K40m GPU with peak performance of 4290 giga floating-point operations per second (GFlop/s) for single-precision. Our code is compiled with compute capability 3.5. As we aim at direct convolution, we compare our kernels with the GEMM-based convolution provided by cuDNN~\cite{cudnn_2014} (version 5.1).


\subsection{Results of Special Case}


\begin{figure}[t]
  \centering
  \includegraphics[width=.85\columnwidth]{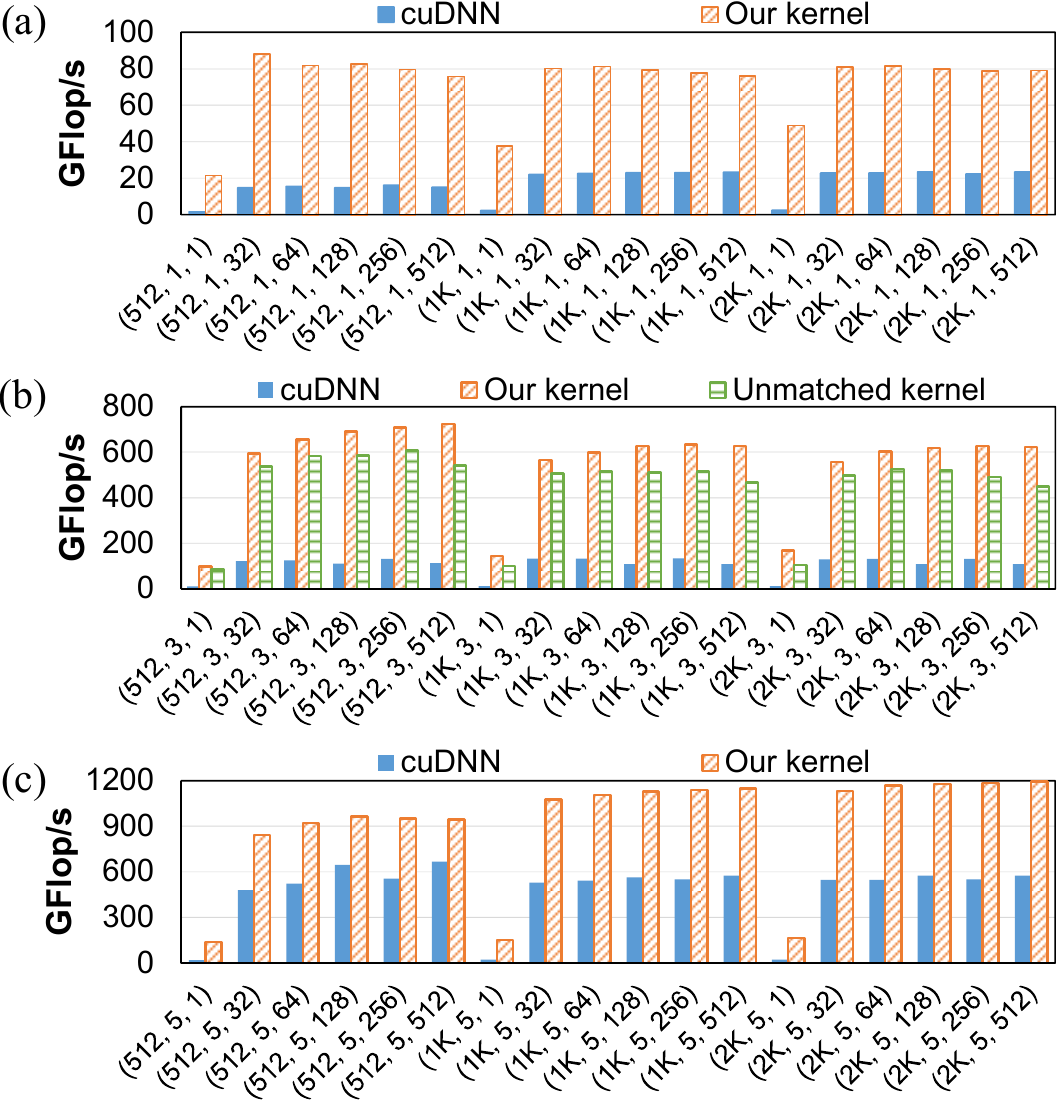}\\
  \vspace{-10pt}
  \caption{Performance of the special case convolution for different convolution parameters $(N,K,F)$. (a) 1$\times$1 filter. (b) 3$\times$3 filter. (c) 5$\times$5 filter.}\label{fig:spec}
   \vspace{-8pt}
\end{figure}

Through design space exploration, we determined that the best block size for the special case convolution kernel is $W\!=\!256$ and $H\!=\!8$. The performance comparison between our kernel and cuDNN for different convolution parameters (image size $N$, filter size $K$, and number of filters $F$) is shown in Fig.~\ref{fig:spec}. For the 1$\times$1 filter, actually there is no data sharing; however, our kernel still obtains an average 6.16$\times$ performance gain, due to the well-designed communication-optimal kernel. For the 3$\times$3 and 5$\times$5 filters, our kernel obtains 6.43$\times$ and 2.90$\times$ average performance gains over cuDNN, respectively. The average performance gain of the three filters we have tested is 5.16$\times$. The performance is lower when $F\!=\!1$, due to the low overlap between communication and computation, as the computation workload is quite low for $F\!=\!1$. However, our kernel can be more than 10$\times$ faster than cuDNN when $F\!=\!1$.

For the 3$\times$3 filter, we have also implemented another kernel in which \dw and \bw are unmatched, i.e., the basic unit for computation is {\tt float}. As seen from Fig.~\ref{fig:spec}b, the performance is reduced by 19\% if \dw and \bw are unmatched. 
It can be expected that the performance degradation will be higher for the general case if \dw and \bw are unmatched, as the SM is used to store both the input image and the filters in the general case.
If we compare the unmatched kernel with cuDNN, even if \dw and \bw are unmatched, our parallelization strategy is still much better than cuDNN for the special case.

\begin{table}[t]
  \centering
  \scriptsize
  \caption{\small Best configurations of our general case convolution kernel for different filter sizes for Kepler K40m.\vspace{-8pt}}\label{tab:cfg}
  \begin{tabular}{cccc}
  \hline
  {Filter size}&{3$\times$3}&{5$\times$5}&{7$\times$7}\\
  \hline
  {$W$}&{32}&{32}&{64}\\
  {$H$}&{4}&{8}&{4}\\
  {$F_{TB}$}&{64}&{32}&{32}\\
  {$W_T$}&{16}&{8}&{8}\\
  {$F_T$}&{4}&{8}&{8}\\
  {$C_{SH}$}&{2}&{1}&{1}\\
  \hline
  \end{tabular}
\end{table}


\begin{figure}[t]
  \centering
  \includegraphics[width=.98\columnwidth]{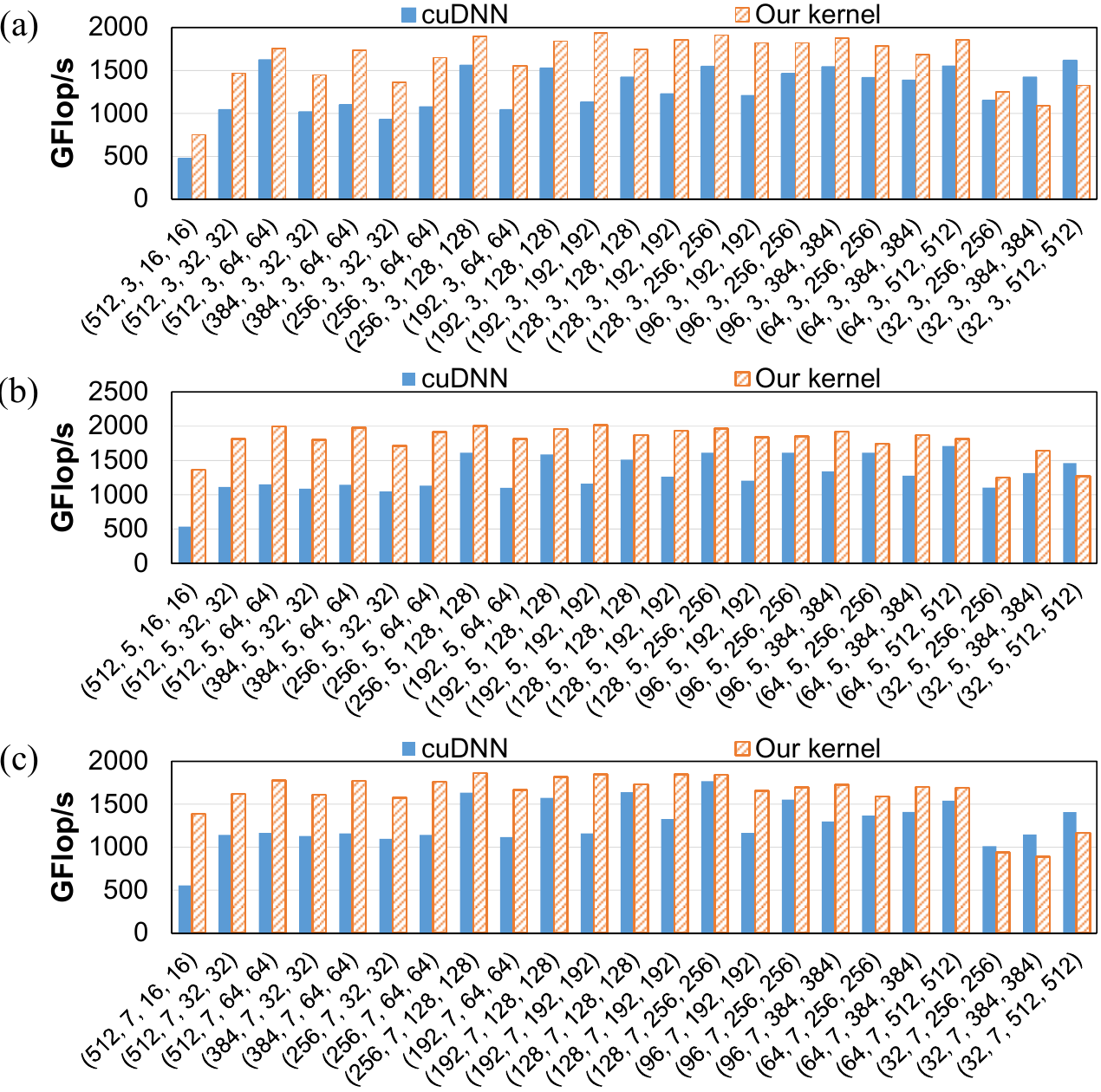}\\
  \vspace{-8pt}
  \caption{Performance of the general case convolution for different convolution parameters $(N,K,C,F)$. (a) 3$\times$3 filter. (b) 5$\times$5 filter. (c) 7$\times$7 filter.}\label{fig:genr}
  \vspace{-8pt}
\end{figure}

\subsection{Results of General Case}
Table~\ref{tab:cfg} lists the best configurations of our general case convolution kernel for different filter sizes for the Kepler K40m GPU. The performance comparison between our kernel and cuDNN for different convolution parameters (image size $N$, filter size $K$, number of channels $C$, and number of filters $F$) is shown in Fig.~\ref{fig:genr}. For the three filter sizes we tested, we get 30.5\%, 45.3\%, and 30.8\% average improvements over cuDNN. Only when the image is very small (32$\times$32), our kernel may be a little slower than cuDNN. In all the other cases, our kernel is always faster than cuDNN. The average performance improvement of the three filter sizes is 35.5\%. The highest performance we have achieved is 2020 GFlop/s, which is 47\% of the hardware peak performance.

\section{Conclusions}\label{sec:concl}
In this paper, we introduced a general model to address the mismatch between the SM bank width and computation data width of threads. Based on this model, we designed and optimized two convolution kernels on GPUs. By carefully optimizing the thread layout and memory access patterns, we attained 5.16$\times$ and 35.5\% average performance improvements over the latest cuDNN library, for the special case and the general case, respectively.

Although we have only implemented our convolution kernels on the Kepler architecture, our proposed ideas can be applied to other applications and architectures. For example, one of the recent development trends of CNNs is to use shorter data types, such as half-precision floating-points and 16- or 8-bit fixed-points, to reduce both the storage requirement and execution time. For these data types, mismatch between the SM bank width and the computation data width exists even for architectures with 4-byte SM bank width. As a result, our proposed model and method will benefit applications using these data types.


\vspace{-8pt}
\small
\section{Acknowledgments}
This project was supported by the National Science Foundation under grants 1640081, 1217906, 1629914 and 1617735, and the Nanoelectronics Research Corporation (NERC), a wholly owned subsidiary of the Semiconductor Research Corporation (SRC), through Extremely Energy Efficient Collective Electronics (EXCEL), an SRC-NRI Nanoelectronics Research Initiative under Research Task IDs 2698.004 and 2698.005.

\vspace{-6pt}
\scriptsize
\bibliographystyle{unsrt}
\bibliography{a}

\begin{thebibliography}{10}

\bibitem{image_book2007}
Rafael~C Gonzalez and Richard~E Woods.
\newblock {\em {Digital Image Processing}}.
\newblock Pearson, 3rd edition, 2007.

\bibitem{image_tmi1989}
S.~Chaudhuri, S.~Chatterjee, N.~Katz, M.~Nelson, and M.~Goldbaum.
\newblock {Detection of Blood Vessels in Retinal Images Using Two-dimensional
  Matched Filters}.
\newblock {\em IEEE Transactions on Medical Imaging}, 8(3):263--269, 1989.

\bibitem{cnn_1989}
Y.~LeCun, B.~Boser, J.~S. Denker, D.~Henderson, R.~E. Howard, W.~Hubbard, and
  L.~D. Jackel.
\newblock {Backpropagation Applied to Handwritten Zip Code Recognition}.
\newblock {\em Neural Comput.}, 1(4):541--551, 1989.

\bibitem{cnn_2014}
Karen Simonyan and Andrew Zisserman.
\newblock {Very Deep Convolutional Networks for Large-Scale Image Recognition}.
\newblock {\em CoRR}, abs/1409.1556, 2014.

\bibitem{cnn_nips2012}
Alex Krizhevsky, Ilya Sutskever, and Geoffrey~E Hinton.
\newblock {Imagenet Classification with Deep convolutional Neural Networks}.
\newblock In {\em NIPS}, pages 1097--1105, 2012.

\bibitem{cnn_cvpr2014}
Ross Girshick, Jeff Donahue, Trevor Darrell, and Jitendra Malik.
\newblock {Rich Feature Hierarchies for Accurate Object Detection and Semantic
  Segmentation}.
\newblock In {\em CVPR}, pages 580--587, 2014.

\bibitem{gemm_2006}
Kumar Chellapilla, Sidd Puri, and Patrice Simard.
\newblock {High Performance Convolutional Neural Networks for Document
  Processing}.
\newblock In {\em ICFHR}, 2006.

\bibitem{cudnn_2014}
Sharan Chetlur, Cliff Woolley, Philippe Vandermersch, Jonathan Cohen, John
  Tran, Bryan Catanzaro, and Evan Shelhamer.
\newblock {cuDNN: Efficient Primitives for Deep Learning}.
\newblock {\em CoRR}, abs/1410.0759, 2014.

\bibitem{direct_hpcc2015}
S.~Li, Y.~Zhang, C.~Xiang, and L.~Shi.
\newblock {Fast Convolution Operations on Many-Core Architectures}.
\newblock In {\em HPCC}, pages 316--323, 2015.

\bibitem{direct_2014}
Ben Van~Werkhoven, Jason Maassen, Henri~E. Bal, and Frank~J. Seinstra.
\newblock {Optimizing Convolution Operations on GPUs Using Adaptive Tiling}.
\newblock {\em Future Gener. Comput. Syst.}, 30:14--26, 2014.

\bibitem{convnet2}
{cuda-convnet2. Url: https://code.google.com/archive/p/cuda-convnet2/}.

\bibitem{fft_2013}
Micha{\"{e}}l Mathieu, Mikael Henaff, and Yann LeCun.
\newblock {Fast Training of Convolutional Networks through FFTs}.
\newblock {\em CoRR}, abs/1312.5851, 2013.

\bibitem{fft_2014}
Nicolas Vasilache, Jeff Johnson, Micha{\"{e}}l Mathieu, Soumith Chintala,
  Serkan Piantino, and Yann LeCun.
\newblock {Fast Convolutional Nets With fbfft: A GPU Performance Evaluation}.
\newblock {\em CoRR}, abs/1412.7580, 2014.

\bibitem{fft_2016}
Tyler Highlander and Andres Rodriguez.
\newblock {Very Efficient Training of Convolutional Neural Networks using Fast
  Fourier Transform and Overlap-and-Add}.
\newblock {\em CoRR}, abs/1601.06815, 2016.

\bibitem{winograd_cvpr2016}
Andrew Lavin and Scott Gray.
\newblock {Fast Algorithms for Convolu- tional Neural Networks}.
\newblock In {\em CVPR}, pages 4013--4021, 2016.

\bibitem{winograd_codes2016}
Hyunsun Park, Dongyoung Kim, Junwhan Ahn, and Sungjoo Yoo.
\newblock {Zero and Data Reuse-aware Fast Convolution for Deep Neural Networks
  on GPU}.
\newblock In {\em CODES}, pages 33:1--33:10, 2016.

\bibitem{cublas_web}
{cuBLAS. Url: http://docs.nvidia.com/cuda/cublas/}.

\bibitem{caffe}
Yangqing Jia, Evan Shelhamer, Jeff Donahue, Sergey Karayev, Jonathan Long,
  Ross~B. Girshick, Sergio Guadarrama, and Trevor Darrell.
\newblock {Caffe: Convolutional Architecture for Fast Feature Embedding}.
\newblock {\em CoRR}, abs/1408.5093, 2014.

\bibitem{magma_jhpca2010}
Rajib Nath, Stanimire Tomov, and Jack Dongarra.
\newblock {An Improved MAGMA Gemm For Fermi Graphics Processing Units}.
\newblock {\em Int. J. High Perform. Comput. Appl.}, 24(4):511--515, 2010.

\end{thebibliography}

\end{document}